\documentclass[floatfix,showpacs,twocolumn,nofootinbib,aps,prd]{revtex4}

\pdfoutput=1

\usepackage{graphicx}
\usepackage{hyperref}
\usepackage{amsmath,amsfonts,amssymb,amsthm}

\begin{document}

\title{Flow in Cyclic Cosmology}
\author{William H. Kinney}
\email{whkinney@buffalo.edu}
\author{Azadeh Moradinezhad Dizgah}
\email{am248@buffalo.edu}
\affiliation{Dept. of Physics, University at Buffalo, The State University of New York, Buffalo, NY 14260-1500}

\begin{abstract}
In this paper, we use a known duality between expanding and contracting cosmologies to construct a dual of the inflationary flow hierarchy applicable to contracting cosmologies such as Ekpyrotic and Cyclic models. We show that the inflationary flow equations are invariant under the duality and therefore apply equally well to inflation or to cyclic cosmology. We construct a self-consistent small-parameter approximation dual to the slow-roll approximation in inflation, and calculate the power spectrum of perturbations in this limit. We also recover the matter-dominated contracting solution of Wands, and the recently proposed Adiabatic Ekpyrosis solution.
\end{abstract}


\maketitle

\section{Introduction}
Inflation \cite{Guth:1980zm,Linde:1981mu,Albrecht:1982wi} has been the most successful theory in explaining the early stages of the universe. It provides a simple explanation for why the universe is homogeneous and geometrically flat, solving the well-known horizon and flatness problems of the standard Big Bang. In addition, it provides natural dynamics to produce the primordial perturbations which are the the source of large-scale structure in the universe. The form of the primordial perturbations predicted by inflation are in good agreement with the observations of Cosmic Microwave Background (CMB) anisotropies \cite{Kinney:2008wy,Komatsu:2010fb,Finelli:2009bs}. Despite its success, there are still unanswered questions about inflation, such as the issue of initial conditions, which have motivated the search for alternatives to inflation. The related Ekpyrotic and Cyclic scenarios \cite{Steinhardt:2001st, Khoury:2001wf, Khoury:2001bz, Khoury:2001zk, Khoury:2003rt, Khoury:2004xi, Buchbinder:2007ad, Buchbinder:2007tw, Lehners:2007ac, Lehners:2008vx, Khoury:2009my} are two of the most widely discussed candidates for alternatives to inflation. Both of these models require a phase of slow contraction occuring before a bounce to an expanding phase. Other alternative proposals include String Gas cosmology \cite{Brandenberger:1988aj,Battefeld:2005av}, which postulates an early cosmological loitering phase, and Tachyacoustic cosmology \cite{ArmendarizPicon:2006if,Magueijo:2008pm,Bessada:2009ns}, which is a model with varying speed of sound. 
                                                                                                                                                            
In inflationary cosmology, perturbations are generated inside the Hubble volume at early times. As the universe goes through the phase of accelerated expansion, the comoving horizon shrinks and the perturbations stretch to super-horizon scales and ``freeze out'' to become classical perturbations. During the subsequent matter/radiation dominated era, as the Hubble Horizon grows, the perturbations fall back into the horizon. In contracting models such as Cyclic and Ekpyrotic cosmologies, perturbations are produced during a phase of contraction. The similar prediction of the two dynamically different models is based on the similar feature of both models, namely a {\it shrinking comoving horizon}. In inflation a shrinking comoving horizon is generated by a period of accelerated expansion whereas in cyclic and ekpyrotic models it is generated by a phase of quasi-static contraction. In both cases, quantum perturbations that were once smaller than the horizon size exit the horizon and become classical. There is in fact a duality relating the inflationary and contracting cases which leaves the comoving horizon invariant. Therefore, for every inflationary model which predicts a nearly scale-invariant spectrum of perturbations, there exists a contracting model with the same prediction for perturbations. 

This duality between contracting and expanding scenarios suggests that methods of studying the inflationary paradigm such as the inflationary flow formalism \cite{Hoffman:2000ue,Kinney:2002qn} can be applied to contracting scenarios as well. The inflationary flow hierarchy is a convenient framework for surveying the parametric space of inflation in a model-independent manner. After reviewing flow methods in inflation, we derive a dual to the inflationary flow hierarchy for contracting scenarios and show that the flow equations are invariant under the duality relating the two cases, and therefore can be employed to study the predictions of contracting scenarios in a model-independent way. After studying the background evolution we continue the discussion by investigating the cosmological perturbations in a contracting phase. 

The outline of the paper is as follows: In Sec. \ref{sec:Duality} we review the known duality between the contracting and expanding cosmologies, in Sec. \ref{sec:dualflow}, after a short review of the inflationary flow hierarchy, we derive the dual flow hierarchy for the contracting models and show that the flow equations are invariant under this duality. Section \ref{sec:perturbations} is the study of cosmological perturbations in the case of a contracting universe. After deriving the general mode equations for perturbations, we solve the scalar and tensor mode equations for two approximate cases of dual analogs to power-law inflation and slow-roll inflation for the contracting case. We also show that the recently proposed ``Adiabatic Ekpyrosis'' scenario \cite{Khoury:2009my} is a special case of family of solutions to the dual flow equations. We conclude in section \ref{sec:conclusions}.

\section{Duality between expanding and contracting cosmologies}
\label{sec:Duality}

A number of dualities have been proposed connecting expanding and contracting cosmologies. A duality proposed by Wands links an expanding inflationary universe with equation of state $w = -1$ to a contracting matter-dominated universe, with equation of state $w = 0$ \cite{Wands:1998yp}. In both cases, a scale-invariant spectrum of curvature perturbations is generated. However, such contracting cosmologies are known to be unstable \cite{Gratton:2003pe,Tolley:2007nq}. Another duality proposed by Brustein {\it et al.} \cite{Brustein:1998kq} is based on the effective action for cosmological perturbations, and likewise connects a stable expanding phase to an unstable contracting phase. 
In this paper, we concentrate on a third duality between expanding and contracting Friedmann-Robertson-Walker (FRW) spaces first proposed by Boyle, {\it et al.} \cite{Boyle:2004gv,Piao:2004uq} and extended to non-constant equation of state by Lidsey \cite{Lidsey:2004xd}. Under this duality, which exchanges the scale factor and Hubble parameter, the inflationary flow equations remain invariant, which we show in Sec. \ref{sec:dualflow}. This property makes this duality especially useful, since it allows definition of a self-consistent dual analog to the slow roll approximation in inflation, and calculation of the corresponding perturbation power spectra.

Consider first the case of an expanding FRW space, with zero curvature and metric
\begin{equation}
ds^2 = dt^2 - a^2(t) d {\bf x}^2 = a^2\left(\tau\right) \left(d\tau^2 - d {\bf x}^2\right),
\end{equation}
where $t$ is the coordinate time and the conformal time is defined by $d \tau = dt / a$. The scale factor is increasing with time, so Hubble parameter in the expanding case is positive and determined by the Friedmann Equation
\begin{equation}
H = \left(\frac{\dot a}{a}\right) = +\sqrt{\frac{1}{3 M_P^2} \rho},
\end{equation}
where $M_P \equiv m_{\rm Pl} / \sqrt{8 \pi}$ is the reduced Planck mass, and $\rho$ is the energy density of the cosmological fluid. 
The time dependence of the scale factor is determined by the equation of state $p = w \rho$ of the cosmological fluid via the Raychaudhuri Equation
\begin{equation}
\frac{\ddot a}{a} = - \frac{1}{6 M_P^2} (1 + 3 w) \rho.
\end{equation}
In an expanding FRW space, $w > -1/3$ corresponds to decelerating expansion, $\ddot a < 0$, and $w < -1/3$ corresponds to accelerating expansion (aka inflation), $\ddot a > 0$. From the Friedmann and Raychaudhuri Equations, we have
\begin{equation}
\label{eq:lHla}
\frac{a}{H} \frac{d H}{d a} = - \frac{3}{2} \left(1 + w\right).
\end{equation}
The size of the cosmological horizon is set by the Hubble length $H^{-1}$, and in comoving units is given by
\begin{equation}
\label{eq:comovinghorizon}
d_{\rm H} = (a H)^{-1}.
\end{equation}From Eq. (\ref{eq:lHla}), the comoving horizon evolves as
\begin{equation}
\frac{d (a H)^{-1}}{d a} = \frac{1}{2 a^2 H} \left(1 + 3 w\right).
\end{equation}
Therefore in the case of decelerating expansion, the comoving horizon grows with time,
\begin{equation}
\frac{d (a H)^{-1}}{d a} > 0,\ w > -1/3,
\end{equation}
and in the case of accelerating expansion, the comoving horizon shrinks,
\begin{equation}
\frac{d (a H)^{-1}}{d a} < 0,\ w < -1/3.
\end{equation}
For a shrinking comoving horizon, quantum modes with constant comoving wavenumber $k$ are stretched to superhorizon scales and ``freeze'' as classical perturbations. As the equation of state approaches $w = -1$, the generated density perturbations approach a scale-invariant power spectrum. 

We wish to construct a contracting cosmology which shares with inflation the property that the comoving horizon size shrinks with time, resulting in the generation of a spectrum of primordial superhorizon perturbations. In a contracting universe, the scale factor is decreasing with time, and the Hubble parameter is negative,
\begin{equation}
H = \left(\frac{\dot a}{a}\right) = -\sqrt{\frac{1}{3 M_P^2} \rho}.
\end{equation}
In the contracting case, the condition (\ref{eq:lHla}) is unchanged, which suggests a duality between expanding and contracting universes which exchanges the scale factor $a$ and the expansion rate $H$ 
\begin{eqnarray}
{\tilde a(\phi)} = H(\phi),\cr
{\tilde H(\phi)} = a(\phi),
\end{eqnarray}
so that the comoving horizon (\ref{eq:comovinghorizon}) is invariant,
\begin{equation}
\left({\tilde a}  {\tilde H}\right)^{-1} = \left(a H\right)^{-1}.
\end{equation}
Here a tilde denotes a quantity in the contracting cosmology. Notice that the dual scenario is itself an expanding phase since $\tilde{H}(\phi) = a(\phi) > 0 $, however time reversal results in a contracting phase. The equation of state transforms under the duality as
\begin{equation}
\frac{\tilde a}{\tilde H} \frac{d {\tilde H}}{d \tilde a} = \frac{H}{a} \frac{d a}{d H},
\end{equation}
or
\begin{equation}
{\tilde w} = - \frac{5 + 9 w}{9 \left(1 + w\right)}.
\end{equation}
The comoving horizon in the contracting universe then evolves as
\begin{equation}
\frac{d (a H)^{-1}}{d a} > 0,\ {\tilde w} < -1/3,
\end{equation}
and
\begin{equation}
\frac{d (a H)^{-1}}{d a} < 0,\ {\tilde w} > -1/3.
\end{equation}
The contracting dual of inflation ($w < -1/3$) is then ${\tilde w} > -1/3$. This is the duality of Boyle, {\it et al.}, who showed that for ${\tilde w} = {\rm const.}$, the primordial density perturbations produced in the contracting dual universe are identical to those produced in the corresponding inflationary universe. (Note that the contracting dual of the de Sitter limit of inflation, with $w = -1$, is the case of {\it divergent} pressure, ${\tilde w} \rightarrow \infty$.) 

In the next section we apply the above duality to general inflationary spacetimes, and show that there is a well-defined dual analog of the slow roll hierarchy for inflation.

\section{A flow hierarchy for contracting FRW spaces}
\label{sec:dualflow}

In this section we define a dual analog for the inflationary flow equations \cite{Hoffman:2000ue,Kinney:2002qn}, which make it possible to investigate cosmological dynamics in a model-independent manner.
After a short review of the method of inflationary flow we show that the flow equations are 
invariant under the duality between the contracting and expanding scenarios and therefore can be applied equally well to the case of cyclic cosmology.

Consider a flat FRW space with energy density dominated by a single scalar field $\phi$ with Lagrangian
\begin{equation}
{\mathcal L} = \frac{1}{2} g^{\mu\nu} \partial_\mu \phi \partial_\nu \phi - V\left(\phi\right).
\end{equation}
The equations of motion for the spacetime are given by the Friedmann and Raychaudhuri Equations, which for a homogeneous field $\phi$ are:
\begin{eqnarray}
\label{eq:Friedmann}
H^2 = \left(\frac{\dot a}{a}\right)^2 &=& \frac{1}{3 M_P^2}\left[\frac{1}{2} \dot\phi^2 + V\left(\phi\right)\right],\cr
\left(\frac{\ddot a}{a}\right) &=& - \frac{1}{3M_P^2} \left[\dot\phi^2 - V\left(\phi\right)\right]. 
\end{eqnarray}
The equation of motion for the field $\phi$ is:
\begin{equation}
\label{eq:phieom}
\ddot \phi + 3 H \dot\phi + V'\left(\phi\right) = 0.
\end{equation}
If the field evolution is monotonic in time, we can write the scale factor $a\left(\phi\right)$ and Hubble parameter $H\left(\phi\right)$ as functions of the field $\phi$ rather than time. Equations (\ref{eq:Friedmann}) and (\ref{eq:phieom}) can then be re-written exactly in the Hamilton-Jacobi form
\begin{eqnarray}
\label{eq:Hamjacobi}
V\left(\phi\right) &=& 3 M_P^2 H^2\left(\phi\right) \left[1 - \frac{2 M_P^2}{3} \left(\frac{H'\left(\phi\right)}{H\left(\phi\right)}\right)^2\right],\cr
\dot\phi &=& - 2 M_P^2 H'\left(\phi\right).
\end{eqnarray} 
We define an infinite hierarchy of ``Hubble slow roll parameters'' \cite{Copeland:1993jj,Liddle:1994dx} by taking successive derivatives of the Hubble parameter $H$ with respect to the field $\phi$:
\begin{eqnarray}
\label{eq:definflationparamshierarchy}
\epsilon &\equiv& {2 M_{P}^2 } \left(\frac{H'(\phi)}{H(\phi)}\right)^2, \cr 
\eta &\equiv& {2 M_{P}^2 } \frac{H''(\phi)}{H(\phi)}, \cr
\xi ^2  &\equiv& {4 M_{P}^4}  \frac{H'(\phi) H'''(\phi)}{H^2(\phi)}, \cr 
\vdots \cr   
{}^ \ell \lambda_H &\equiv& {\left(2 M_{P}^2\right) ^ \ell} \frac{H'(\phi)^{\left(\ell-1\right)}}{H(\phi)^ \ell} \frac{d^{\left(\ell +1\right)} H(\phi)}{d \phi ^ {\left(\ell +1\right)}},
\end{eqnarray}
where the prime denotes derivatives with respect to scalar field $\phi$. We refer to this hierarchy as {\it flow parameters}. The first flow parameter $\epsilon$ is related to the equation of state by:
\begin{equation}
\epsilon = \frac{3}{2} \left(1 + w\right) = -\frac{a}{H} \frac{d H}{d a},
\end{equation}
where we have used Eq. (\ref{eq:lHla}). In an expanding universe, the comoving horizon shrinks for $w < -1/3$, or $\epsilon < 1$, which is just the case of inflation. For an inflating universe, the scale factor increases quasi-exponentially, with the Hubble parameter $H \simeq {\rm const.}$ and
\begin{equation}
a \propto \exp\left[\int{H dt}\right] \equiv e^{-N},
\end{equation}
where $N$ is the number of e-folds. The sign convention for $N$ above means that $N$ {\it decreases} with increasing time, and therefore measures the number of e-folds before the end of inflation. The number of e-folds $N$ is related to the field $\phi$ by
\begin{eqnarray}
\label{numberofefolds}
d N \equiv - H dt &=& -\frac{d a(\phi)}{a(\phi)} \cr
 &=& \frac{1}{\sqrt{2} M_P}\frac{d\phi}{\sqrt{\epsilon(\phi)}},
\end{eqnarray}
where we choose the sign convention that $\sqrt{\epsilon}$ has the same sign as $H'\left(\phi\right)$,
\begin{equation}
\sqrt{\epsilon} \equiv + \sqrt{2} M_P \frac{H'\left(\phi\right)}{H\left(\phi\right)}.
\end{equation}
We can then write the parameter $\epsilon$ as:
\begin{equation}
\epsilon = \frac{1}{H} \frac{d H}{d N}.
\end{equation}
Similarly, taking derivatives of the flow parameters with respect to $N$ we can generate an infinite set of differential equations relating the parameters:
\begin{eqnarray} 
\frac{d \epsilon}{d N} &=& 2 \epsilon \left(\eta - \epsilon\right), \cr
\frac{d \eta}{d N} &=&  \xi^2- \epsilon \eta, \cr
\vdots \cr  
\frac{d {}^\ell \lambda_H} {d N} &=& \left[(\ell-1) \eta - \ell \epsilon \right] {}^ \ell \lambda_H + {} ^ {(\ell + 1)} \lambda_H,
\end{eqnarray}
Taken to infinite order, this hierarchy of flow equations completely specifies the evolution of the spacetime \cite{Kinney:2002qn}: once the parameters $\epsilon\left(N\right)$, $\eta\left(N\right)$ and so forth are known, the cosmological dynamics and scalar field potential are fixed via the Hamilton-Jacobi Equations (\ref{eq:Hamjacobi}). 

We wish to construct a similar hierarchy of flow equations applicable to contracting universes, utilizing the duality discussed in Section \ref{sec:Duality}. Under the duality $a \leftrightarrow H$, the parameter $\epsilon$ transforms as:
\begin{equation}
\epsilon = 2 M_P^2 \left(\frac{H'\left(\phi\right)}{H\left(\phi\right)}\right)^2 \rightarrow {\tilde \epsilon} = 2 M_P^2 \left(\frac{a'\left(\phi\right)}{a\left(\phi\right)}\right)^2.
\end{equation}
From the definition of Hubble parameter we have:
\begin{equation}
\frac{H'(\phi)}{H(\phi)} = -\frac{1}{2 M_P^2} \frac{a(\phi)}{a'(\phi)},
\end{equation}
so that the dual parameter $\tilde \epsilon$ can be seen to be the inverse of the flow parameter $\epsilon$,
\begin{equation}
\label{eq:epsinv}
{\tilde \epsilon} = \frac{1}{2 M_P^2} \left(\frac{H(\phi)}{H'(\phi)}\right)^2 = \frac{1}{\epsilon}.
\end{equation}
Under $a \leftrightarrow H$, the dual of the number of e-folds $d N = - da / a$ is:
\begin{eqnarray}
\label{eq:dualN}
d \tilde{N} \equiv - \frac{d H(\phi)}{H(\phi)} &=& -\frac{H'(\phi)}{H(\phi)}d\phi \cr
 &=& \frac{1}{\sqrt{2} M_P} \frac{d \phi}{\sqrt{{\tilde \epsilon}(\phi)}},
\end{eqnarray}
where we choose the sign convention that $\sqrt{\tilde \epsilon}$ has the same sign as $a'\left(\phi\right)$ and therefore the {\it opposite} sign as $H'\left(\phi\right) / H\left(\phi\right)$,
\begin{equation}
\sqrt{\tilde \epsilon} \equiv + \sqrt{2} M_P \frac{a'\left(\phi\right)}{a\left(\phi\right)} = -\frac{1}{2 M_p^2} \frac{H\left(\phi\right)}{H'\left(\phi\right)}.
\end{equation}
The number of e-folds and its dual are related by:
\begin{equation}
d N = - {\tilde\epsilon} d \tilde{N},
\end{equation}
so that we can write $\epsilon$ and ${\tilde \epsilon}$ as:
\begin{equation}
\epsilon = - \frac{d {\tilde N}}{d N},\qquad \qquad {\tilde \epsilon} = - \frac{d N}{d {\tilde N}}.
\end{equation}
Defining ${\mathcal N}$ as the logarithm of the comoving horizon size,
\begin{equation}
{\mathcal N} \equiv - \ln{\left(a H\right)},
\end{equation}
we see that ${\mathcal N}$ is invariant under the duality $a \leftrightarrow H$, and is related to $N$ and ${\tilde N}$ by:\
\begin{equation}
d {\mathcal N} = \left(1 - \epsilon\right) dN = \left(1 - {\tilde \epsilon}\right) d {\tilde N},
\end{equation}
In an inflating universe, the comoving horizon shrinks ($d {\mathcal N} / dN > 0$) for $\epsilon < 1$. Likewise, in a contracting universe, the comoving horizon shrinks for $\tilde \epsilon < 1$, so that $\tilde\epsilon$ plays the same role in contracting universes that $\epsilon$ plays in expanding universes. We define the full hierarchy of dual flow parameters by replacing $H\left(\phi\right)$ in Eqs.(\ref{eq:definflationparamshierarchy}) with $a\left(\phi\right)$,  
\begin{eqnarray}
\label{eq:defcontractingparamshierarchy}
{\tilde\epsilon} &\equiv& {2 M_{P}^2 } \left(\frac{a'(\phi)}{a(\phi)}\right)^2, \cr 
\tilde{\eta} &\equiv& {2 M_{P}^2 } \frac{a''(\phi)}{a(\phi)}, \cr
\tilde{\xi} ^2  &\equiv& {4 M_{P}^4}  \frac{a'(\phi) a'''(\phi)}{a^2(\phi)}, \cr 
\vdots \cr   
{}^ \ell \tilde{\lambda}_a &\equiv& {\left(2 M_{P}^2\right) ^ \ell} \frac{a'(\phi)^{\left(\ell-1\right)}}{a(\phi)^ \ell} \frac{d^{\left(\ell +1\right)} a(\phi)}{d \phi ^ {\left(\ell +1\right)}},
\end{eqnarray}
Comparing Eqs. (\ref{numberofefolds}) and (\ref{eq:dualN}), we see that the dual $d {\tilde N}$ has exactly the same form as $d N$, so that the dual flow equations are identical to their counterparts in inflation:
\begin{eqnarray}
\label{contractingflow} 
\frac{d {\tilde\epsilon}}{d \tilde{N}} &=& 2 {\tilde\epsilon} \left(\tilde{\eta} - {\tilde\epsilon}\right), \cr
\frac{d \tilde{\eta}}{d \tilde{N}} &=& \tilde{\xi}^2-{\tilde\epsilon}\tilde{\eta}, \cr
\vdots \cr  
\frac{d {}^\ell \tilde{\lambda}_a}{d \tilde{N}} &=& \left[(\ell-1) \tilde{\eta} - \ell {\tilde\epsilon}\right] {}^ \ell \tilde{\lambda}_a + {} ^ {(\ell + 1)}\tilde{\lambda}_a.
\end{eqnarray}
This duality invariance of the flow equations means that for every inflationary potential, there exists an equivalent dual cyclic potential for which the comoving horizon has the same dynamics \cite{Boyle:2004gv}.

We can then define a dual analog of the slow-roll limit of inflation by taking ${\tilde \epsilon} \ll 1$ and ${\tilde \eta} \ll 1$, so that 
\begin{equation}
\frac{d \epsilon}{d {\tilde N}} \ll \epsilon,
\end{equation}
and so forth, with higher order flow parameters small. The flow parameters can then be expressed approximately in terms of the potential by using the Hamilton-Jacobi Equations,
\begin{eqnarray}
\label{eq:Hamjacobidual}
V\left(\phi\right) &=& 3 M_P^2 H^2\left(\phi\right) \left(1 - \frac{1}{3 {\tilde \epsilon}\left(\phi\right)}\right), \cr
\dot\phi &=& - 2 M_P^2 H'\left(\phi\right).
\end{eqnarray}
For the dual-slow roll limit $\tilde\epsilon \ll 1$,
\begin{eqnarray}
\label{eq:VdualSR}
V\left(\phi\right) &\simeq& - M_P^2 \frac{H^2\left(\phi\right)}{{\tilde \epsilon}\left(\phi\right)}\cr
&=& - 2 M_P^4 \left[H'\left(\phi\right)\right]^2 = -\frac{1}{2} \dot\phi^2.
\end{eqnarray}
In the case of slow roll inflation, $\epsilon \propto \left(H'/H\right)^2 \ll 1$ means that the Hubble parameter $H$ and scale factor $a$ evolve as:
\begin{eqnarray}
H^2 &\simeq& \frac{1}{3 M_P^2} V\left(\phi\right) \simeq {\rm const.},\cr
a &\sim& e^{H t}.
\end{eqnarray}
In the dual limit ${\tilde \epsilon} \propto \left(a'/a\right)^2$, we have
\begin{eqnarray}
a &\simeq& {\rm const.},\cr
H^2 &=& \frac{1}{3 M_P^2} \left[\frac{1}{2}\dot\phi^2 + V\left(\phi\right)\right] \simeq 0.
\end{eqnarray}
The dual limit of slow roll inflation is therefore a quasi-static contraction, with the scale factor nearly time-invariant. Note that the potential in the cyclic dual to inflation must be negative.  Differentiating Eq. (\ref{eq:VdualSR}) with respect to $\phi$ gives
\begin{eqnarray}
\frac{V'\left(\phi\right)}{V\left(\phi\right)} &\simeq& 2 \frac{H'}{H} - \frac{{\tilde \epsilon}'}{\tilde \epsilon}\cr
&=& 2 \frac{H'}{H} \left(1 + {\tilde \eta} - {\tilde \epsilon}\right),
\end{eqnarray}
where we have used the flow equations (\ref{contractingflow}) and
\begin{equation}
\frac{d}{d \phi} = \frac{1}{M_P \sqrt{2 {\tilde \epsilon}}} \frac{d}{d {\tilde N}} = - \frac{H'}{H} \frac{d} {d {\tilde N}}.
\end{equation}
Then, to lowest order in the flow parameters,
\begin{equation}
\tilde\epsilon = \frac{1}{2 M_P^2} \left(\frac{H\left(\phi\right)}{H'\left(\phi\right)}\right)^2 \simeq \frac{2}{M_P^2} \left(\frac{V\left(\phi\right)}{V'\left(\phi\right)}\right)^2.
\end{equation}
Using the first flow equation (\ref{contractingflow}), the second flow parameter $\tilde\eta$ can be written in the dual-slow roll limit as:
\begin{eqnarray}
{\tilde \eta} &=& {\tilde \epsilon} + \frac{1}{2 {\tilde \epsilon}} \frac{d {\tilde \epsilon}}{d {\tilde N}}\cr
&\simeq& -2 \left[1 - \frac{1}{M_P^2} \left(\frac{V}{V'}\right)^2 - \frac{V V''}{\left(V'\right)^2}\right].
\end{eqnarray}
This differs slightly from the ``fast roll'' parameter $\eta$ defined in Ref. \cite{Khoury:2003rt}, but is physically equivalent. We hesitate to adopt the terminology ``fast roll'' because the dual limit of slow roll inflation is not the limit of $\dot\phi^2 \gg V\left(\phi\right)$, but the highly tuned circumstance that the kinetic and potential contributions to the energy density almost exactly cancel. We will use the term {\it dual-slow roll} to describe this situation. 

In this section, we have defined a dual flow hierarchy which is self-consistent to arbitrary order and which admits a dual analog of the slow roll approximation, applicable to cyclic cosmology. The dual version of the quasi-exponential expansion of inflation is a quasi-static contraction, with nearly constant scale factor. In both cases, the comoving horizon size is shrinking with time. The flow equations governing the evolution of the spacetime are {\it invariant} under the duality transformation. In the next section, we derive equations for cosmological perturbations in the case of a contracting universe.

\section{Cosmological perturbations}
\label{sec:perturbations}

\subsection{Scalar power spectra}
\label{sec:generalperturbations}

The scalar perturbations in a flat FRW cosmology with a single scalar field can be expressed in terms of a single gauge-invariant parameter. Two commonly used parameters are the Bardeen potential $\Phi$, corresponding to the potential in Newtonian gauge, and the curvature 
perturbation $\zeta$, corresponding to the curvature perturbation in comoving gauge. Assuming for simplicity $p = w \rho$, with $w = {\rm const.}$, the gauge-independent variables $\zeta$ and $\Phi$ are not independent, but are related by:
\begin{equation}
\label{zeta and phi}
\zeta = \frac{2}{3 a^2 (1+w)} \left(\frac{\Phi}{a'/a^3}\right)',
\end{equation}
where a prime denotes a derivative with respect to conformal time. Ultimately, the physical variable which is most directly related to CMB perturbations is the Newtonian potential, {\it i.e.} the Bardeen potential $\Phi$. In the case of inflation, it is conventional to express the perturbations in terms of $\zeta$ since it is conserved on superhorizon scales even when the equation of state is varying, which is not true for $\Phi$. 

Let us consider the relation between the two gauge-invariant potentials in more detail. For simplicity we again take the case of constant equation of state, $w = const.$  For a flat FRW cosmology, the Friedmann and Raychaudhuri equations are written as:
\begin{eqnarray}
\left(\frac{a'}{a ^ 2}\right)^2 &=& \frac{1}{3 M_P^2} \rho, \cr 
\frac{a''}{a} &=& - \frac{1}{6 M_P^2} (1 + 3w) \rho,
\end{eqnarray}
where prime again denotes derivatives with respect to conformal time $\tau$. Therefore, the scale factor $a$ evolves as
\begin{equation}
\label{scalefactor}
a \propto \tau^{2 / 3(1 + w)}.
\end{equation}
There are two sets of expanding and contracting solutions:
\begin{itemize}
\item Expanding: 
\begin{eqnarray}
w &>& -1/3,  \quad \tau \in [0, \infty] ,\cr
w &<& -1/3,  \quad \tau \in [- \infty,0].
\end{eqnarray} 
\item Contracting:
\begin{eqnarray}
w &>& -1/3,  \quad \tau \in [-\infty, 0] ,\cr
w &<& -1/3,  \quad \tau \in [0, \infty].
\end{eqnarray}
\end{itemize}
The equation of motion for the Bardeen potential $\Phi$ is:
\begin{equation}
{\Phi''} + 3(1 + w)( aH ) {\Phi'} + wk^2\Phi = 0.
\end{equation}
Using
\begin{equation}
a H = \frac{2}{1 + 3w} \tau ^{-1},
\end{equation}
we have
\begin{equation}
{\Phi''} + \frac{6(1 + w)}{1 + 3w} \tau^{-1} {\Phi'} + wk^2 \Phi = 0,
\end{equation}
which in the long wavelength limit, $k \rightarrow 0 $, reduces to
\begin{equation}
{\Phi''} + \frac{6(1 + w)}{1 + 3w} \tau ^{-1} {\Phi'} = 0.
\end{equation}
This equation has two solutions:
\begin{equation}
\label{eq:solutions}
\Phi_0 = const. ,  \qquad \qquad \Phi_1 = \tau ^{ - (5 + 3w)/(1 + 3w)}.
\end{equation}
Note that in an expanding universe, $\Phi_1$ is always a decaying mode, regardless of $w$, since
\begin{equation}
 w > -1/3,  \quad \tau \in [0, \infty]  \ \Longrightarrow   \ - \frac{5 + 3w}{1 + 3w} < 0 , \nonumber
\end{equation}
\begin{equation}
\qquad  \ w < -1/3,  \quad \tau \in [-\infty, 0] \  \Longrightarrow  \ - \frac{5 + 3w}{1 + 3w} > 0.
\end{equation}
In the case of inflation, $w < -1/3$ and $\tau$ is negative and tends towards zero. Therefore $\Phi_1$ is again a decaying mode. Therefore, in an expanding cosmology, the physically relevant mode is the constant mode $\Phi_0$. Using Eq. (\ref{scalefactor}), the transformation (\ref{zeta and phi}) becomes,
\begin{equation}
\zeta = \frac{2 \left[\Phi' + \left(a H\right)\Phi\right]}{3 H \left(1 + w\right)} + \Phi,
\end{equation}
so that for $\Phi = \Phi_0 = {\rm const.}$, $\zeta$ and $\Phi$ have the simple relationship
\begin{equation}
\zeta = \frac{5 + 3 w}{3 \left(1 + w\right)} \Phi.
\end{equation}
Since $\zeta$ is conserved {\it even when $w$ is changing}, we can calculate $\zeta$ during inflation, and then calculate the Bardeen potential $\Phi_f$ in any later epoch with equation of state $w_f$ as \cite{Mukhanov:2005sc}:
\begin{equation}
\Phi_f = \frac{3 \left(1 + w_f\right)}{5 + 3 w_f} \zeta.
\end{equation}
This is the standard lore for perturbations in inflation. 

The solution $\Phi_1$ behaves quite differently in the case of a contracting universe;
\begin{equation}
w > -1/3,  \quad \tau \in [-\infty, 0]  \ \Longrightarrow   \ - \frac{5 + 3w}{1 + 3w} < 0 , \nonumber
\end{equation}
\begin{equation}
\qquad  \ w < -1/3,  \quad \tau \in [0, \infty] \  \Longrightarrow  \ - \frac{5 + 3w}{1 + 3w} > 0.
\end{equation}
Therefore from Eq. (\ref{eq:solutions}) $\Phi_1$ is always a {\it growing} mode irrespective of $w$ and  is the mode of physical interest. Using Eq. (\ref{scalefactor}), we have
\begin{equation}
\Phi_1 \propto \frac{a'}{a^3}.
\end{equation}
From Eq. (\ref{zeta and phi}) it is then apparent that $\Phi_1$ does not contribute to the curvature perturbation $\zeta$. Therefore, in a contracting universe, the gauge-invariant variable $\zeta$ tracks {\it only} the subdominant mode of the Newtonian potential \cite{Gratton:2003pe,Boyle:2004gv}. In a contracting universe, we must also follow the Bardeen potential $\Phi$.
It is more convenient to work with two new parameters $u$ and $v$ which are related to $\Phi$ and $\zeta$ by \cite{Mukhanov:1988jd}:
\begin{equation}
u \equiv {\left(\frac{2 M_p^2 a}{\phi'}\right) \Phi},  \quad \qquad  v \equiv {z \zeta}, 
\end{equation}
where z is given by:
\begin{equation}
\label{z}
z = \frac{\sqrt{2} M_P a}{\sqrt{\tilde\epsilon}}.
\end{equation}
The variables $u$ and $v$ are not independent, but are related by:
\begin{equation}
\label{transformationvu}
u = - \frac{1}{k^2} z \left(\frac{v}{z}\right)', 
\end{equation}
\begin{equation}
\label{transformationuv}
v = \theta \left(\frac{u}{\theta}\right)'.
\end{equation}
These transformation equations are equivalent to independent mode equations for $u$ and $v$:
\begin{eqnarray}
\label{modequation}
v_k'' + \left(k^2 - \frac{z''}{z}\right)v_k = 0, \cr
u_k'' + \left( k^2 -\frac{\theta ''}{\theta}\right)u_k = 0,
\end{eqnarray}
where $\theta = 1/z$, $k$ is the conformal wave number, and a prime denotes derivative with respect to conformal time $d \tau \equiv dt/a$. We choose to 
work with a dimensionless parameter $y$ instead of $\tau$, defined as:
\begin{equation}
y \equiv \frac{k}{a H}.
\end{equation} 
Writing
\begin{equation}
\frac{d}{d\tau}=a\frac{d}{dt}=-aH \frac{d}{d N},
\end{equation}
we obtain a differential relation between $y$ and $\tau$,
\begin{equation}
dy=\left(\frac{1-{\tilde\epsilon}}{{\tilde\epsilon}}\right)k d \tau,
\end{equation}
so for the second derivative we have
\begin{equation}
\label{derivative}
\frac{d^2}{d \tau ^2} = \frac{y^2}{\tau ^2}\frac{d^2}{dy^2} - \frac{2 y \left(\tilde{\eta} - {\tilde\epsilon}\right)}{\left(1- {\tilde\epsilon}\right)^2 \tau ^2} \frac{d}{dy}.
\end{equation}

From Eq. (\ref{z}) and using contracting flow equations (\ref{contractingflow}) we obtain $\theta''/\theta$ and $z''/z$  in terms of flow 
parameters
\begin{eqnarray}
\label{z and theta}
\frac{z''}{z} &&= \left(\frac{aH}{{\tilde\epsilon}}\right)^2 F \left({\tilde\epsilon}, \tilde{\eta},\tilde{\xi}^2\right), \cr 
\frac{\theta''}{\theta} &&= \left(\frac{aH}{{\tilde\epsilon}}\right)^2 G \left({\tilde\epsilon}\tilde{\eta}, \tilde{\xi}^2\right),
\end{eqnarray}

where
\begin{eqnarray}
F \left({\tilde\epsilon}, \tilde{\eta},\tilde{\xi}^2\right) && \equiv \left[-\tilde{\xi}^2+ 3 \tilde{\eta}^2+ 6 {\tilde\epsilon}^2 - 6 {\tilde\epsilon}\tilde{\eta}+\tilde{\eta} -2{\tilde\epsilon}\right], \cr 
G \left({\tilde\epsilon}, \tilde{\eta},\tilde{\xi}^2\right) &&\equiv \left[\tilde{\xi} ^2 - \tilde{\eta}^2 + 2 {\tilde\epsilon}^2 - 2 \tilde{\eta}{\tilde\epsilon}-\tilde{\eta} + 2 {\tilde\epsilon}\right].
\end{eqnarray}

Substituting (\ref{z and theta}) and (\ref{derivative}) into (\ref{modequation}) we find
\begin{align}
\label{generalmode}
 \left(1-\tilde{\epsilon}\right)^2 y ^2 \frac{d^2 v_k}{dy^2} &- 2y\left(\tilde{\eta}-{\tilde\epsilon}\right)\frac{dv_k}{dy} \nonumber \\
\qquad \qquad  &+ \left[y^2 {\tilde\epsilon}^2 - F \left({\tilde\epsilon}, \tilde{\eta},\tilde{\xi}^2\right)\right]v_k = 0, 
\end{align}
\begin{align}
\left(1- \tilde{\epsilon}\right)^2 y ^2 \frac{d^2 u_k}{dy^2} &- 2y\left(\tilde{\eta}-{\tilde\epsilon}\right)\frac{du_k}{dy} \nonumber \\
\qquad \qquad &+\left[y^2 {\tilde\epsilon}^2 - G \left({\tilde\epsilon}, \tilde{\eta},\tilde{\xi}^2\right)\right]u_k = 0,
\end{align}
which are exact mode equations for $u$ and $v$. In the following sections we solve the mode equations in three approximate limits: power law contraction, the dual-slow roll limit, and Adiabatic Ekpyrosis. In the power-law case, we also recover the matter-dominated contracting solution of Wands.

\subsection{Power Law Case}

In this section, we consider the dual analog to power-law inflation, which has the useful property that it is possible to solve exactly for both the background evolution and the perturbation. Power-law inflation corresponds to taking the first flow parameter $\epsilon$ to be exactly constant,
\begin{equation}
\epsilon = \frac{1}{H}\frac{d H}{d N} = {\rm const.},
\end{equation}
so that the Hubble parameter and scale factor evolve as:
\begin{equation}
H \propto e^{\epsilon N},\qquad a \propto e^{-N}.
\end{equation}
The comoving horizon then evolves as:
\begin{equation}
\left(a H\right)^{-1} \propto e^{\left(1 - \epsilon\right) N},
\end{equation}
which shrinks for $\epsilon < 1$. It is straightforward to show that the scale factor evolves as a power law in time,
\begin{equation}
\label{eq:PLinf}
a \propto t^{1 / \epsilon}.
\end{equation}
The dual of power-law inflation for the contracting case corresponds to taking the flow parameter $\tilde\epsilon$ to be exactly constant,
\begin{equation}
\label{eq:PLeps}
{\tilde \epsilon} = \frac{1}{a}\frac{d a}{d {\tilde N}} = {\rm const.},
\end{equation}
so that the Hubble parameter and scale factor evolve as:
\begin{equation}
a \propto e^{{\tilde \epsilon} {\tilde N}},\qquad H \propto e^{-{\tilde N}}.
\end{equation}
The comoving horizon then evolves as:
\begin{equation}
\left(a H\right)^{-1} \propto e^{\left(1 - {\tilde \epsilon}\right) {\tilde N}},
\end{equation}
which shrinks for ${\tilde \epsilon} < 1$. Eq. (\ref{eq:PLeps}) can be simply solved for the scale factor as a function of the field $\phi$,
\begin{equation}
a\left(\phi\right) \propto \exp\left(\sqrt{\frac{\tilde\epsilon}{2}} \frac{\phi}{M_P}\right).
\end{equation}
Likewise, from Eq. (\ref{eq:epsinv}), we have
\begin{equation}
H\left(\phi\right) = H_0 \exp\left(-\frac{1}{\sqrt{2 {\tilde\epsilon}}} \frac{\phi}{M_P}\right),
\end{equation} 
so that the scalar field potential can be written using the Hamilton-Jacobi equation (\ref{eq:Hamjacobidual}),
\begin{equation}
V\left(\phi\right) = \left(\frac{3 {\tilde\epsilon} - 1}{\tilde\epsilon}\right) M_P^2 H_0^2 \exp\left(-\sqrt{\frac{2}{\tilde\epsilon}} \frac{\phi}{M_p}\right).
\end{equation}
As in the case of inflation, the scale factor is a power law in time, and is the dual of Eq. (\ref{eq:PLinf}),
\begin{equation}
a \propto t^{\tilde\epsilon}.
\end{equation}

To express the mode equations (\ref{generalmode}) in the power-law case, note that for ${\tilde\epsilon} = {\rm const.}$, the flow equations (\ref{contractingflow}) are solved exactly by expressing higher-order flow parameters as powers of $\tilde\epsilon$,
\begin{equation}
{\tilde\epsilon} = {\rm const.}  , \qquad  \tilde{\eta} = {\tilde\epsilon}, \qquad \tilde{\xi}^2 = {\tilde\epsilon}^2, \qquad {}^\ell{\tilde\lambda} = {\tilde\epsilon}^\ell,
\end{equation}
so the mode equations for $v$ and $u$ are reduced to
\begin{eqnarray}
\left(1-\tilde{\epsilon}\right)^2 y^2 \frac{d^2 v_k}{dy^2}+ \left[{\tilde\epsilon}^2 y^2 + {\tilde\epsilon}\left(1 - 2 {\tilde\epsilon}\right)\right]v_k = 0, \cr
\left(1-{\tilde\epsilon}\right)^2 y^2 \frac{d^2 u_k}{dy^2}+ \left[{\tilde\epsilon}^2 y^2 -{\tilde\epsilon}\right]u_k = 0,
\end{eqnarray}
with solutions
\begin{eqnarray}
\label{eq:vusolutions}
v_k(y) = \sqrt{y} \left[c_1 H^{(1)}_\beta \left(\frac{{\tilde\epsilon}y}{{\tilde\epsilon} - 1}\right) + c_2 H^{(2)}_\beta \left(\frac{{\tilde\epsilon}y}{{\tilde\epsilon}-1}\right)\right], \cr
u_k(y) = \sqrt{y} \left[d_1 H^{(1)}_\alpha \left(\frac{{\tilde\epsilon}y}{{\tilde\epsilon} - 1}\right) + d_2 H^{(2)}_\alpha \left(\frac{{\tilde\epsilon}y}{{\tilde\epsilon}-1}\right)\right],
\end{eqnarray}  
where $\alpha$ and $\beta$ are given by:
\begin{eqnarray}
\beta &\equiv& \frac{1}{2}\left|\frac{1 - 3 {\tilde\epsilon}}{1 - {\tilde\epsilon}}\right|, \cr
\alpha &\equiv& \frac{1}{2}\left|\frac{1+ {\tilde\epsilon}}{1 - {\tilde\epsilon}}\right|.
\end{eqnarray}
The integration constants are fixed by specifying appropriate boundary conditions. As in the case of inflation, the relevant boundary conditions correspond to canonical quantization of the perturbations and the selection of a vacuum state, which we take to be the usual Bunch-Davies vacuum. For a detailed discussion of the quantization of the functions $v$ and $u$, see Ref. \cite{Mukhanov:1990me}: $v$ and $u$ are in fact canonical conjugates of one another, and the canonically quantized variable is $v$. Once $v$ has been fixed by quantization and vacuum selection, the function $u$ (which corresponds to the Bardeen potential $\Phi$) is determined by applying the transformation (\ref{transformationvu}).

The asymptotic form of (\ref{eq:vusolutions}) for the short wavelength limit, ${\tilde\epsilon} y \rightarrow \infty $, is given by: 
\begin{equation}
v_k(y) = \sqrt{\frac{2}{\pi}}\left( c_1 e^{-ik\tau} +c_2 e^{+ik \tau}\right).
\end{equation}
Selecting the Bunch-Davies vacuum as the vacuum state of fluctuations in the ultraviolet limit,
\begin{equation}
v_k \propto e^{-ik\tau},
\end{equation} 
requires that $c_2=0$. Canonical quantization fixes the other constant $c_1$. The canonical commutation relation for $v$ is
\begin{equation}
\left[v\left({\bf x} , \tau \right), \pi\left({\bf x'},\tau\right) \right] = i \delta ^3 \left({\bf x}- {\bf x'}\right), 
\end{equation}
which corresponds to a Wronskian condition for the mode $v_k$,
\begin{equation}
v_k \frac{\partial v^*_k}{\partial \tau} - v^*_k \frac{\partial v_k}{\partial \tau} = i.
\end{equation}
The fully normalized solution for $v$ is given by
\begin{equation}
\label{eq:vnormPL}
v_k(y) = \frac{1}{2}\sqrt{\frac{\pi}{k}\left(\frac {{\tilde\epsilon}y}{1-{\tilde\epsilon}}\right)}H^{(1)}_\beta \left(\frac{{\tilde\epsilon}y}{{\tilde\epsilon}-1}\right),
\end{equation}
which agrees with the result of \cite{Boyle:2004gv}. Having the solution for $v$ we use the exact transformation relation between $u$ and $v$ (\ref{transformationvu}) to obtain $u$,
\begin{eqnarray}
\label{uinpowerlaw}
u(y) &=& - \frac{1}{k^2}\left[v' - v \frac{z'}{z}\right] \cr 
     &=& -\frac{1}{k^2}\left[\frac{dv}{dy}\frac{dy}{d\tau} - v \left(\frac{z'}{z}\right)\right].
\end{eqnarray}
In the power law limit ${\tilde\epsilon} = \tilde{\eta} $, so we have
\begin{eqnarray}
\frac{z'}{z} &=& \frac{k}{y {\tilde\epsilon}}\left(2 {\tilde\epsilon} - \tilde{\eta}\right) \cr
             &=& \frac{k}{y},
\end{eqnarray}
and (\ref{uinpowerlaw}) reduces to
\begin{equation}
\label{powerlawtransform}
u_k(y) = \frac{1}{k} \left[\frac{v}{y}-\left(\frac{1-{\tilde\epsilon}}{\tilde\epsilon}\right)\frac{dv}{dy}\right].
\end{equation}
Therefore the fully normalized solution for $u$ is:
\begin{equation}
\label{eq:unormPL}
u_k(y) = - \frac{1}{2}\sqrt{\frac{\pi}{k^3}\left(\frac {{\tilde\epsilon}y}{1-{\tilde\epsilon}}\right)}H^{(1)}_\alpha \left(\frac{{\tilde\epsilon}y}{{\tilde\epsilon}-1}\right).
\end{equation}
Note that the argument of the Bessel function is ${\tilde\epsilon} y / \left(1 - {\tilde \epsilon}\right) \simeq {\tilde\epsilon} y$. Therefore, mode freezing occurs when
\begin{equation}
{\tilde\epsilon y} = \frac{{\tilde\epsilon} k}{a H} \simeq 1,
\end{equation}
so that in a contracting universe with ${\tilde\epsilon} \ll 1$, quantum modes freeze out and become classical well inside the Hubble length, $k \gg \left(a H\right)$.
The power spectra for the curvature perturbation $\zeta$ and for the Newtonian potential $\Phi$ are obtained by taking the long-wavelength limit,
\begin{eqnarray}
\label{eq:PLzetaspec}
&&P_\zeta(k) = \frac{k^3}{2 \pi^2} \left|\frac{v_k}{z}\right|_{{\tilde\epsilon} y \rightarrow 0}^2 \cr
&&= \frac{2^{2\beta -4}}{M_P^2}\left(\frac{\Gamma(\beta)}{\Gamma(3/2)}\right)^2 {\tilde\epsilon}\left(\frac{{\tilde\epsilon}}{1 - {\tilde\epsilon}}\right)^{1 - 2\beta}\left(\frac{H}{2\pi}\right)^2 y^{3-2\beta}, \nonumber \\
\end{eqnarray}
and
\begin{eqnarray}
\label{eq:PLphispec}
&&P_\Phi(k)  = \frac{k^3}{2 \pi^2} \left|\frac{H z}{2 M_p^2 a}  u_k\right|_{{\tilde\epsilon} y \rightarrow 0}^2 \cr
&&= \frac{2^{2\alpha -2}}{4 M_P^2} \left(\frac{\Gamma(\alpha)}{\Gamma(3/2)}\right)^2 \left(\frac{{\tilde\epsilon}}{1 - {\tilde\epsilon}}\right)^{1 - 2\alpha}\frac{1}{{\tilde\epsilon}}\left(\frac{H}{2\pi}\right)^2 y^{1-2\alpha}. \nonumber \\
\end{eqnarray}
The spectral index is then calculated by differentiating the above equations with respect to $k$ at constant time, $aH=const.$ \cite{Kinney:2005vj},
\begin{equation}
\label{eq:PLscalarn}
n_\zeta -1 = \left|\frac{d \ln P_\zeta}{d \ln k}\right|_{aH=const.}
           = 3 - 2 \beta = \frac{2}{1-{\tilde\epsilon}},
\end{equation}
and
\begin{equation}
n_\Phi -1 = \left|\frac{d \ln P_\Phi}{d \ln k}\right|_{aH=const.} 
           = 1 - 2 \alpha = \frac{- 2 {\tilde\epsilon}}{1-{\tilde\epsilon}}.
\end{equation}
In the limit $\tilde\epsilon \rightarrow 0$, the power spectrum for $\Phi$ approaches scale invariance, while the power-spectrum for $\zeta$ approaches the well-known result of a strongly blue spectrum $n_\zeta = 3$. For this case to represent a viable cosmological model, we must have mode mixing at the bounce, {\it i.e.} the growing mode for $\Phi$ in the contracting phase must map to the constant mode in the expanding phase. The scale-invariant solution of Wands \cite{Wands:1998yp} is simply the limit $\tilde\epsilon = 2/3$, which results in $n_\zeta = 1$, which requires no mode mixing at the bounce.

In the next section, we solve for the perturbations in the limit dual to the slow-roll limit of inflation. 
 
\subsection{Dual-slow roll limit}

The power-law case discussed in the previous section has ${\tilde\epsilon} = {\tilde\eta}$, and allows exact solution of the background and perturbation equations. This is the dual analog of the power-law limit of inflation, with $\epsilon = \eta$. The slow roll limit of inflation is distinct, with $\epsilon$ and $\eta$ independent, but both small. In this section, we consider the dual of the slow roll limit, such that the dual parameters are independent and small:
\begin{equation}
{\tilde\epsilon} \neq {\tilde \eta},   \qquad {\tilde\epsilon} ,\tilde{\eta} \ll 1 .
\end{equation}
Solving the mode equations in the dual-slow roll limit is more involved. Unlike the power law case where the term proportional to first order derivative of the mode function vanishes since $\tilde {\epsilon} = \tilde{\eta}$, in the dual-slow roll limit this term persists and therefore the Bunch-Davies vacuum is not the vacuum state for equation (\ref{generalmode}). For this term to vanish, and to obtain a consistent mode equation, we use a new parameter $x$ which is related to the  dimensionless parameter $y$ by:
\begin{equation}
\frac{d}{dy}={\tilde\epsilon} \frac{d}{dx}.
\end{equation}
Therefore there is an identical relation between this new dimensionless parameter $x$ and $\tau$ as that between $y$ and $\tau$,
\begin{eqnarray}
k d \tau &=& \left(1- \epsilon\right) dy, \cr
k d \tau &=& \left(1-{\tilde\epsilon}\right) dx.
\end{eqnarray}
 
The second derivatives are related by:
\begin{equation}
\frac{d^2}{dy^2} = \frac{2 {\tilde\epsilon} \left({\tilde\eta}-{\tilde\epsilon}\right)}{y\left(1-{\tilde\epsilon}\right)}\frac{d}{dx} + {\tilde\epsilon}^2 \frac{d^2}{dx^2}.
\end{equation}
Hence the mode equation in terms of this new variable is:
\begin{eqnarray}
\label{eq:modex}
\left({\tilde\epsilon}y\right)^2 \left(1-{\tilde\epsilon}\right)^2
\frac{d^2v_k}{dx^2} &+& 2\left({\tilde\epsilon}y\right)\left[{\tilde
\epsilon}\left({\tilde
\epsilon}-\tilde{\eta}\right)\right]\frac{dv_k}{dx} \cr
&+& \left[\left({\tilde\epsilon} y\right)^2+F\left({\tilde\epsilon},\tilde{\eta},
\tilde{\xi}^2\right)\right]v_k=0.
 \nonumber \\
\end{eqnarray}
To have a well-formulated Bessel equation we need to write the variable ${\tilde\epsilon}y$ in the coefficients in terms of $x$. We have a differential 
relation between the two variables. We can expand $x$ as an infinite series in flow parameters by integrating by parts,
\begin{eqnarray}
\label{eq:xseries}
x &&= \int{{\tilde\epsilon} dy} = {\tilde\epsilon}y- \int{y \frac{d{\tilde\epsilon}}{dy} dy} \cr
  &&= {\tilde\epsilon}y+ \frac{2 {\tilde\epsilon}\left({\tilde\epsilon}-\tilde{\eta}\right)}{1-{\tilde\epsilon}}y + {\mathcal O} \left({\tilde\epsilon}^3, {\tilde\epsilon}^2 {\tilde\eta} \right)+\hdots  \ . 
\end{eqnarray} 
Therefore, to first order in the flow parameters,
\begin{equation}
\label{eq:xSR}
x \simeq {\tilde\epsilon}y.
\end{equation}
We then get a dual-slow roll mode equation,
\begin{equation}
x^2\left(1-2{\tilde\epsilon}\right) \frac{d^2v}{dx^2} + \left(x^2 + 2 {\tilde\epsilon}-\tilde{\eta}\right)v = 0,
\end{equation}
with solution
\begin{equation}
v_k(x) \longrightarrow \sqrt{x} \left[c_1 H^{(1)}_{\beta}\left(\frac{-x}{\sqrt{1-2{\tilde\epsilon}}}\right)+c_2 H^{(2)}_{\beta}\left(\frac{-x}{\sqrt{1-2{\tilde\epsilon}}}\right)\right],
\end{equation}
where 
\begin{equation}
\label{betaslowroll}
\beta = \frac{\sqrt{1-10 {\tilde\epsilon} + 4\tilde{\eta}}}{2\sqrt{1-2{\tilde\epsilon}}} \simeq \frac{1}{2}-2 {\tilde\epsilon}+\tilde{\eta}
\end{equation}
Similar to the power law case, the choice of Bunch-Davies vacuum as the vacuum state of fluctuations and canonical quantization of $v$ gives the fully normalized solution,
\begin{equation}
v_k(y) = \frac{1}{2}\sqrt{\frac{\pi}{k}\left(\frac{{\tilde\epsilon}y}{1-{\tilde\epsilon}}\right)}H^{(1)}_\beta \left(\frac{{\tilde\epsilon}y}{{\tilde\epsilon}-1}\right).
\end{equation}

Once again the solution for $v$ can be placed in (\ref{transformationvu}) to find $u$. The transformation from $v$ to $u$ in this case is
\begin{equation}
\label{slowrolltransform}
u(y) = \frac{1}{k} \left[\frac{2 {\tilde\epsilon}- \tilde{\eta}}{y {\tilde\epsilon}}v -\left(\frac{1-{\tilde\epsilon}}{\tilde\epsilon}\right)\frac{dv}{dy}\right],
\end{equation}
since
\begin{equation}
\frac{z'}{z} = \frac{k}{y {\tilde\epsilon}}\left(2 {\tilde\epsilon} - \tilde{\eta}\right). 
\end{equation}
The fully normalized solution for $u$ is then
\begin{equation}
u_k(y) = \frac{(-1)^{\alpha}}{2}\sqrt{\frac{\pi}{k^3} \left(\frac{{\tilde\epsilon}y}{1-{\tilde\epsilon}}\right)}H^{(1)}_\alpha \left(\frac{{\tilde\epsilon}y}{{\tilde\epsilon}-1}\right),
\end{equation}
where
\begin{equation}
\label{alphaslowroll}
\alpha = \frac{\sqrt{1 + 6{\tilde\epsilon} - 4\tilde{\eta}}}{2\sqrt{1-2{\tilde\epsilon}}} \simeq \frac{1}{2}+ 2 {\tilde\epsilon} - \tilde{\eta}.
\end{equation}

The equations for the power spectra for $\zeta$ and $\Phi$ are the same as in the power law case (\ref{eq:PLzetaspec},\ref{eq:PLphispec}) except that the indices $\alpha$ and $\beta$ are different and are given by (\ref{betaslowroll}) and (\ref{alphaslowroll}). The corresponding spectral indices are:
\begin{equation}
n_\zeta -1 = \left|\frac{d \ln P_\zeta}{d \ln k}\right|_{aH=const.} 
           = 3 - 2 \beta = 2 + 4 {\tilde\epsilon}-2\tilde{\eta}, 
\end{equation}
and
\begin{equation}
n_\Phi -1 = \left|\frac{d \ln P_\Phi}{d \ln k}\right|_{aH=const.} 
           = 1 - 2 \alpha = -4{\tilde\epsilon} + 2\tilde{\eta}.
\end{equation}
This can be compared to the result for the case of inflation,
\begin{equation}
n_\Phi - 1 = n_\zeta - 1 = -4 \epsilon + 2 \eta.
\end{equation}
The power spectrum for the Bardeen potential $\Phi$ is therefore invariant under the duality $\epsilon \rightarrow {\tilde \epsilon}$, $\eta \rightarrow {\tilde \eta}$ connecting inflationary and cyclic cosmology. In the next section, we consider the recently proposed ``Adiabatic Ekpyrosis'' scenario.

\subsection{Adiabatic Ekpyrosis}

Khoury and Steinhardt have recently proposed the {\it Adiabatic Ekpyrosis} scenario \cite{Khoury:2009my}, which is a contracting universe producing a scale-invariant spectrum for the curvature perturbation $\zeta$. In Adiabatic Ekpyrosis, the parameter $\epsilon$ evolves with time as: 
\begin{equation}
\epsilon = \frac{1}{H} \frac{d H}{d N} = \frac{1}{t^2},
\end{equation}
with $a(t)$ approximately constant, so that
\begin{equation}
z \propto a \sqrt{\epsilon} \propto \frac{1}{t} \propto \frac{1}{\tau},
\end{equation}
which results in a scale-invariant power spectrum for $\zeta$. In this section we use the dual flow formalism to generalize this solution to a family of solutions parametrized by the second flow parameter $\tilde\eta$. 

To construct the required class of solutions, note that the flow equations (\ref{eq:defcontractingparamshierarchy}) are solved by the following {\it ansatz}:
\begin{eqnarray}
&&{\tilde\epsilon} = {}^\ell{\tilde\lambda} = 0,\cr
&&{\tilde\eta} = {\rm const.}
\end{eqnarray}
This is the dual of the non-slow-roll inflationary solutions considered in Refs. \cite{Kinney:1997ne,Kinney:2005vj,Tzirakis:2007bf,Tzirakis:2008qy}, and does not require the parameter $\tilde\eta$ to be small, only constant. This is an exact solution for $\tilde\epsilon$ exactly vanishing, and approximate for $\tilde\epsilon \ll 1$. We solve for the quantum mode $v_k$ using the mode equation (\ref{eq:modex}), in terms of the variable $dx \equiv {\tilde\epsilon} dy$. However, for $\tilde\eta = {\rm const.} \sim {\mathcal O}(1)$, the expansion (\ref{eq:xseries}) cannot be truncated after one term as in Eq. (\ref{eq:xSR}). This can be seen by writing the first flow equation to lowest order in ${\tilde\epsilon}$,
\begin{equation}
y \frac{d {\tilde\epsilon}}{d y} = \frac{1}{1 - {\tilde\epsilon}} \frac{d {\tilde\epsilon}}{d {\tilde N}} = \frac{2 {\tilde\epsilon} \left({\tilde \eta} - {\tilde\epsilon}\right)}{1 - {\tilde\epsilon}} \simeq 2 {{\tilde \epsilon}{\tilde \eta}}.
\end{equation}
The series of partial integrations in Eq. (\ref{eq:xseries}) is then an infinite series in $\tilde\eta$,
\begin{equation}
x = \left({\tilde\epsilon} y\right) \sum_{n=0}^{\infty}{\left(- 2 {\tilde \eta}\right)^n} = \frac{{\tilde\epsilon} y}{1 + 2 {\tilde\eta}}.
\end{equation}
The mode equation (\ref{eq:modex}) is then
\begin{equation}
x^2 \frac{d^2 v_k}{d x^2} + \left(x^2 - \frac{{\tilde\eta}\left(1 + 3 {\tilde \eta}\right)}{\left(1 + 2 {\tilde\eta}\right)^2}\right) v_k = 0,
\end{equation}
with normalized solution
\begin{equation}
\label{eq:adiabaticsol}
v_k\left(x\right) = \frac{1}{2}\sqrt{\frac{\pi x}{k}}H^{(1)}_\beta \left(x\right),
\end{equation}
where
\begin{equation}
\label{eq:betaAE}
\beta = \left\vert \frac{1 + 4 {\tilde \eta}}{2 \left(1 + 2 {\tilde \eta}\right)} \right\vert.
\end{equation}
The transformation (\ref{slowrolltransform}) in this limit becomes
\begin{eqnarray}
u_k\left(x\right) &=& - \frac{1}{k} \left[\frac{\tilde\eta}{1 + 2 {\tilde\eta}} \left(\frac{v_k\left(x\right)}{x}\right) + \frac{d v_k\left(x\right)}{d x}\right]\cr
&=& - \frac{1}{2} \sqrt{\frac{\pi x}{k^3}} H^{(1)}_{\alpha}\left(x\right),
\end{eqnarray}
where
\begin{equation}
\label{eq:alphaAE}
\alpha = \left\vert \beta - 1\right\vert = \left\vert \frac{1}{2 \left(1 + 2 {\tilde \eta}\right)}\right\vert.
\end{equation}
The spectral indices for $\zeta$ and $\Phi$ are then, for ${\tilde \eta} < -1/2$,
\begin{equation}
n_\zeta - 1 = 3 - 2 \beta = \frac{2 \left(1 + {\tilde\eta}\right)}{1 + 2 {\tilde\eta}},
\end{equation}
and
\begin{equation}
n_\Phi - 1 = 1 - 2 \alpha = \frac{2 \left(1 + {\tilde\eta}\right)}{1 + 2 {\tilde\eta}}.
\end{equation}
Therefore for ${\tilde\eta} = -1$, the power spectra for $\zeta$ and $\Phi$ are identical and scale-invariant. (This is not in contradiction to the discussion in Sec. \ref{sec:generalperturbations}, since in Adiabatic Ekpyrosis, the equation of state $w$ is changing rapidly.) It is straightforward to show that ${\tilde\eta} = -1$ corresponds to the case $\epsilon = 1 / {\tilde\epsilon} = (1/t^2)$ considered by Khoury and Steinhardt, since
\begin{equation}
\epsilon = - \frac{\dot H}{H^2} = \frac{1}{t^2}\ \Longrightarrow H = \frac{-t}{1 + C t}.
\end{equation}
Therefore, for $t \rightarrow 0$,
\begin{equation}
\frac{d}{d t} = \frac{H}{\tilde\epsilon} \frac{d}{d {\tilde N}} = - \frac{1}{t} \frac{d}{d {\tilde N}}.
\end{equation}
The first flow equation is then
\begin{equation}
\frac{d {\tilde\epsilon}}{d {\tilde N}} = 2 {\tilde\epsilon} \left({\tilde\eta} - {\tilde\epsilon}\right) =  - 2 t^2 = - 2 {\tilde\epsilon},
\end{equation}
so that
\begin{equation}
{\tilde\eta} = {\tilde\epsilon} - 1 \simeq -1,
\end{equation}
and we recover the solution of Khoury and Steinhardt \cite{Khoury:2009my} as a special case of the solution (\ref{eq:adiabaticsol}). For $\left\vert {\tilde\eta}\right\vert \ll 1$, we recover the dual-slow roll expressions
\begin{equation}
n_\zeta - 1 = \frac{2 \left(1 + {\tilde\eta}\right)}{1 + 2 {\tilde\eta}} \simeq 2 - 2 {\tilde\eta},
\end{equation}
and 
\begin{equation}
n_\Phi - 1 =  \frac{2 {\tilde\eta}}{1 + 2 {\tilde\eta}} \simeq 2 {\tilde \eta}.
\end{equation}
Adiabatic Ekpyrosis therefore generates a scale-invariant power spectrum for cosmological perturbations. However, Linde {\it et al.} have pointed out that this scenario suffers from problems of trans-Planckian curvatures and a breakdown of linear perturbation theory \cite{Linde:2009mc}.

In the next section, we consider the generation of tensor perturbations in contracting universes, and show that the tensor spectrum is generically blue, and tensors are strongly suppressed relative to the scale-invariant mode of the scalar spectrum.

\subsection{Tensor perturbations}
In this section, we calculate the spectrum of tensor (gravitational wave) perturbations produced in a contracting universe. To first order, the mode equation for tensor perturbations is:
\begin{equation}
h_k''+\left(k^2-\frac{a''}{a}\right)h_k = 0,
\end{equation}
where
\begin{equation}
\frac{a''}{a} = \left(aH\right)^2\frac{2{\tilde\epsilon}-1}{{\tilde\epsilon}}.
\end{equation}
Again changing the variable from $\tau$ to $y$, the mode equation for tensor perturbations is:
\begin{align}
\left(1-{\tilde\epsilon}\right)^2 y^2 \frac{d^2 h_k}{dy^2} &- 2y \left(\tilde{\eta}-{\tilde\epsilon}\right)\frac{d h_k}{dy} \nonumber \\
\qquad \qquad  &+ \left[y^2 {\tilde\epsilon}^2 - {\tilde\epsilon}\left(2 {\tilde\epsilon}-1\right)\right]h_k = 0.
\end{align}

We first consider the case of power law contraction, ${\tilde \epsilon} = {\tilde \eta} = {\rm const.}$. In this limit, the mode equation is:
\begin{equation}
\left(1-{\tilde\epsilon}\right)^2y^2 \frac{d^2h_k}{dy^2} +\left[y^2{\tilde\epsilon}^2 - 2{\tilde\epsilon}^2 + {\tilde\epsilon}\right]h_k = 0,
\end{equation}
and the normalized solution is given by:
\begin{equation}
\label{eq:hPL}
h_k(y) = \frac{1}{2}\sqrt{\frac{\pi}{k}\left(\frac{{\tilde\epsilon}y}{1-{\tilde\epsilon}}\right)}H^{(1)}_\gamma\left(\frac{{\tilde\epsilon}y}{{\tilde\epsilon}-1}\right),
\end{equation}
where
\begin{equation}
\label{eq:gammaPL}
\gamma = \frac{1}{2}\left|\frac{1-3{\tilde\epsilon}}{1-{\tilde\epsilon}}\right|.
\end{equation}
The power spectrum for tensors in the long-wavelength limit is then
\begin{eqnarray}
&&P_T(k) = \frac{8}{M_P^2} \left(\frac{k^3}{2 \pi^2}\right) \left\vert \frac{h_k}{a}\right\vert_{{\tilde \epsilon} y \rightarrow 0}\cr
&&= \frac{2^{2\gamma}}{M_P^2}\left(\frac{\Gamma(\gamma)}{\Gamma(3/2)}\right)^2 \left(\frac{{\tilde\epsilon}}{1 - {\tilde\epsilon}}\right)^{1 - 2\gamma}\left(\frac{H}{2\pi}\right)^2 y^{3-2\gamma}. \nonumber \\
\end{eqnarray} 
The spectral index for tensor perturbations is
\begin{equation}
n_T = \left|\frac{d \ln P_T}{d \ln k}\right|_{aH=const.} 
           = 3 - 2 \gamma = \frac{2}{1-{\tilde\epsilon}}.
\end{equation}
The tensor spectrum in a contracting universe is therefore strongly blue, with $n_{\rm T} \sim 2$, and has the same spectral index as the curvature perturbation $\zeta$ (\ref{eq:PLscalarn}), $n_{\rm T} = n_\zeta - 1$. This means we can define an analog of the tensor/scalar ratio in inflation, $r_\zeta$, as:
\begin{equation}
r_\zeta \equiv \frac{P_{\rm T}}{P_{\zeta}} = \frac{16}{\tilde \epsilon},
\end{equation}
where we have used Eq. (\ref{eq:PLzetaspec}) for the power spectrum of the curvature perturbation $\zeta$. This is exactly the dual of the inflationary expression $r = 16 \epsilon$. However, in a collapsing universe with ${\tilde\epsilon} \ll 1$, tensors are enhanced relative to the curvature perturbation, rather than suppressed as in the case of inflation. Nevertheless, both the tensor and curvature perturbations are suppressed relative to the Bardeen potential $\Phi$, since
\begin{eqnarray}
\label{tentoscalratio}
r_\Phi &\equiv& \frac{P_{\rm T}}{P_\Phi}\cr
&=& 16 {\tilde\epsilon} \left(\frac{\Gamma\left(\gamma\right)}{\Gamma\left(\alpha\right)}\right)^2 \left(\frac{{\tilde\epsilon}}{2 \left({\tilde\epsilon}-1\right)}\right)^{2 \left( \alpha - \gamma\right)} y^{2 + 2 \alpha - 2 \gamma}. \nonumber \\
\end{eqnarray}
Noting that
\begin{equation}
2 \left(\alpha - \gamma\right) = \frac{4 {\tilde\epsilon}}{1 - {\tilde\epsilon}},
\end{equation}
in the nearly scale-invariant limit $\tilde\epsilon \ll 1$,
\begin{equation}
r_\Phi \simeq 16 \tilde\epsilon y^2 \big\vert_{y \rightarrow 0} \longrightarrow 0.
\end{equation}
This makes physical sense, since both $P_\zeta$ and $P_{\rm T}$ are constant modes on superhorizon scales \cite{Kinney:2005vj}, 
\begin{equation}
H^2 y^{3 - 2 \gamma} = {\rm const.},
\end{equation}
while $P_\Phi$ is a growing mode,
\begin{equation}
H^2 y^{1 - 2 \alpha} \propto y^{-2 - 2\alpha + 2 \gamma}.
\end{equation}
The ratio $r_\zeta$ is therefore constant, but the ratio $r_\Phi$ is negligible on superhorizon scales, $y \ll 1$. 

In dual-slow roll limit, the mode equation is given by:
\begin{equation}
\left(1-2{\tilde\epsilon}\right)x^2 \frac{d^2 h_k}{d x^2} + \left(x^2 + {\tilde\epsilon}\right)h_k = 0.
\end{equation}
The fully normalized solution is then
\begin{equation}
h_k(y) = \frac{1}{2}\sqrt{\frac{\pi}{k} \left(\frac{{\tilde\epsilon}y}{1-{\tilde\epsilon}}\right)}H^{(1)}_\gamma \left(\frac{{\tilde\epsilon}y}{{\tilde\epsilon}-1}\right),
\end{equation}
which is the same as the solution (\ref{eq:hPL}) in the power law case except that the order of Hankel function $\gamma$ is: 
\begin{equation}
\label{gammaslowroll}
\gamma = \frac{\sqrt{1-6{\tilde\epsilon}}}{2\sqrt{1-2{\tilde\epsilon}}} \simeq \frac{1}{2} - {\tilde\epsilon},
\end{equation}
which can be seen as the small-$\tilde\epsilon$ limit of Eq. (\ref{eq:gammaPL}). The tensor spectral index is then
\begin{equation}
n_T -1 = \left|\frac{d \ln P_T}{d \ln k}\right|_{aH=const.} = 3 - 2 \gamma = 2 + 2 {\tilde\epsilon}. 
\end{equation}
The tensor/scalar ratio $r_\Phi$ is given by (\ref{tentoscalratio}) with $\alpha$ and $\gamma$ given by (\ref{alphaslowroll}) and (\ref{gammaslowroll}). Having
\begin{equation}
2\alpha -2 \gamma = 6 {\tilde\epsilon} - 2 \tilde{\eta},
\end{equation}
in the dual-slow roll limit, ${\tilde\epsilon} \ll 1$ and $\tilde{\eta} \ll 1$, the tensor/scalar ratio reduces to an expression identical to the power-law case,
\begin{equation}
r_\Phi \simeq 16{\tilde\epsilon} y^2| _{y \rightarrow 0} \longrightarrow 0.
\end{equation}
In both the power-law and dual-slow roll cases, the tensor spectrum has a suppressed amplitude relative to the Bardeen potential $\Phi$, but has constant amplitude relative to the curvature perturbation $\zeta$. The prediction of cyclic cosmology for the tensor/scalar ratio is therefore highly dependent on the presence or absence of mode mixing in the transition from a contracting universe to an expanding universe. 

Finally, we consider the case of Adiabatic Ekpyrosis. The long-wavelength asymptotic limits for the power spectra in $\zeta$ and $\Phi$ are:
\begin{eqnarray}
&&P_\zeta(k) = \frac{k^3}{2 \pi^2} \left|\frac{v_k}{z}\right|_{{\tilde\epsilon} y \rightarrow 0}^2 \cr
&&= \frac{2^{2\beta -4}}{M_P^2}\left(\frac{\Gamma(\beta)}{\Gamma(3/2)}\right)^2 {\tilde\epsilon}\left(\frac{{\tilde\epsilon}}{1 + 2 {\tilde\eta}}\right)^{1 - 2\beta}\left(\frac{H}{2\pi}\right)^2 y^{3-2\beta}, \nonumber \\
\end{eqnarray}
and
\begin{eqnarray}
&&P_\Phi(k)  = \frac{k^3}{2 \pi^2} \left|\frac{H z}{2 M_p^2 a}  u_k\right|_{{\tilde\epsilon} y \rightarrow 0}^2 \cr
&&= \frac{2^{2\alpha -2}}{4 M_P^2} \left(\frac{\Gamma(\alpha)}{\Gamma(3/2)}\right)^2 \left(\frac{{\tilde\epsilon}}{1 + 2 {\tilde\eta}}\right)^{1 - 2\alpha}\frac{1}{{\tilde\epsilon}}\left(\frac{H}{2\pi}\right)^2 y^{1-2\alpha}. \nonumber \\
\end{eqnarray}
The constants $\beta$ and $\alpha$ are given by Eq. (\ref{eq:betaAE}) and Eq. (\ref{eq:alphaAE}), respectively. The tensor/scalar ratios for $\zeta$ and $\Phi$ are then
\begin{eqnarray}
&&r_\zeta = \frac{P_{\rm T}}{P_\zeta} \cr
&&= \left(\frac{16}{\tilde\epsilon}\right) \left(\frac{\Gamma\left(\gamma\right)}{\Gamma\left(\beta\right)}\right)^2 \left(\frac{\tilde\epsilon}{2}\right)^{2\left(\beta - \gamma\right)} \frac{\left(1 + 2 {\tilde\eta}\right)^{1 - 2 \beta}}{\left(1 - {\tilde\epsilon}\right)^{1 - 2 \gamma}} y^{2 \left(\beta - \gamma\right)},\nonumber \\
\end{eqnarray}
and 
\begin{eqnarray}
&&r_\Phi = \frac{P_{\rm T}}{P_\Phi} \cr
&&= 16 {\tilde\epsilon} \left(\frac{\Gamma\left(\gamma\right)}{\Gamma\left(\alpha\right)}\right)^2 \left(\frac{\tilde\epsilon}{2}\right)^{2\left(\alpha - \gamma\right)} \frac{\left(1 + 2 {\tilde\eta}\right)^{1 - 2 \alpha}}{\left(1 - {\tilde\epsilon}\right)^{1 - 2 \gamma}} y^{2 - 2 \gamma + 2 \alpha}.\nonumber \\
\end{eqnarray}
In the scale-invariant limit, ${\tilde\epsilon} \ll 1$ and ${\tilde\eta} \simeq -1$, the tensor/scalar ratios are
\begin{equation}
r_\zeta = r_\Phi \simeq 16 {\tilde\epsilon} y^{2}| _{y \rightarrow 0} \longrightarrow 0.
\end{equation}
The tensor/scalar ratio is strongly suppressed for both $\zeta$ and $\Phi$.

\section{Conclusions}
\label{sec:conclusions}
It has been shown by several authors \cite{Wands:1998yp, Boyle:2004gv, Lidsey:2004xd, Piao:2004uq} that there exist dualities between the contracting and expanding cosmologies in their predictions for primordial perturbations. In this paper we have considered the duality of Boyle {\it et al.} and Lidsey \cite{Boyle:2004gv, Lidsey:2004xd}, which corresponds to an exchange of the scale factor $a(\phi)$  and Hubble Parameter $H(\phi)$. The comoving Hubble horizon $d_H = (a(\phi)H(\phi))^{-1}$ is invariant under this duality and therefore identical perturbation spectra can be produced in inflating and contracting cosmologies. We have introduced a self-consistent flow hierarchy for contracting cosmologies and have shown that the flow equations are invariant under this duality.   

After studying the evolution of a contracting background in terms of the flow parameters we have investigated the evolution of the cosmological perturbations. The mode equations of perturbations in the gauge-invariant Bardeen potential $\Phi$, corresponding to the Newtoian potential, are second order differential equations, and therefore have two solutions: a constant and growing/decaying mode. In the expanding phase the non-constant mode is decaying and the physically relevant mode is the constant mode. This constant mode of perturbations in the Bardeen potential maps onto the curvature perturbation $\zeta$, so that we can follow $\zeta$, which is conserved on superhorizon scales, to track the perturbations. However, in a corresponding contracting phase, the non-constant mode is a growing mode, which does not contribute to perturbations in $\zeta$. Therefore, in a contracting phase we must follow $\Phi$ as well as $\zeta$ to completely describe cosmological perturbations. Having derived the general mode equations for perturbations in $\zeta$ and $\Phi$, we solved the equations in three approximate limits:  power-law contraction, dual-slow roll, and Adiabatic Ekpyrosis. In the power-law limit our result is in agreement with the existing results in the literature. In the slow-roll limit, our calculation of the spectral indices indicates that the spectral index of Bardeen potential is invariant under this duality since:  
\begin{equation}
n_\Phi -1 = -4 \tilde{\epsilon} + 2 \tilde{\eta},
\end{equation}
which has the same form as the spectral index of the Bardeen potential in inflation:
\begin{equation}
n_\Phi - 1 = - 4 \epsilon + 2 \eta.
\end{equation}
In contrast, $\zeta$ has highly blue spectrum,
\begin{equation}
n_\zeta - 1 = 2 + 4 {\tilde \epsilon} - 2 {\tilde \eta}.
\end{equation}
In the case of Adiabatic Ekpyrosis \cite{Khoury:2009my}, both $\zeta$ and $\Phi$ acquire nearly scale-invariant spectra,
\begin{equation}
n_\zeta - 1 = n_\Phi - 1 = \frac{2 \left(1 + {\tilde\eta}\right)}{1 + 2 {\tilde\eta}},
\end{equation}
where the scale-invariant limit is ${\tilde\eta} = -1$. 

We also defined the analog of the tensor/scalar ratio $r$ in inflation,  for contracting models. We have shown that the tensor/scalar ratio for the curvature perturbation is invariant under the duality,
\begin{equation}
\label{eq:rzeta}
r_\zeta  \simeq \frac{16}{\tilde{\epsilon}}.
\end{equation} 
However for the Bardeen potential in both the power law and dual-slow roll limits, the tensor spectrum is suppressed relative to Bardeen potential,
\begin{equation}
\label{eq:rPhi}
r_\Phi = 16 \tilde{\epsilon} y^2 |_{y \rightarrow 0}  \longrightarrow 0 ,
\end{equation}
where
\begin{equation}
y \equiv \frac{k}{aH}.
\end{equation}
In the case of Adiabatic Ekpyrosis, tensors are suppressed relative to both $\zeta$ and $\Phi$,
\begin{equation}
r_\zeta = r_\Phi \simeq 16 {\tilde\epsilon} y^{2}| _{y \rightarrow 0} \longrightarrow 0.
\end{equation}
What we have calulated in this paper are the spectra of cosmological perturbations in contracting cosmologies {\it before} the bounce and subsequent cosmological expansion. A central issue remains: how do these primordial spectra translate into perturbations {\it after} the bounce? If there is no mode mixing at the bounce, the constant mode $\zeta$ in the contracting phase will match to the corresponding constant mode in the expanding phase, whereas the growing mode $\Phi$ will match to a decaying mode in the expanding phase \cite{Brandenberger:2001bs,Tsujikawa:2002qc,Allen:2004vz,Gasperini:2004ss,Copeland:2006tn,Wands:2008tv}. In such a situation, the mode of physical interest is $\zeta$, and the contracting dual of slow roll inflation predicts a strongly blue power spectrum, inconsistent with the data. However, the matching of modes at the bounce is known to be highly model-dependent \cite{Martin:2001ue, Durrer:2002jn, Peter:2002cn, Hwang:2002ks, Cartier:2003jz, Khoury:2003vb, Tolley:2003nx, Gasperini:2003pb, Peter:2003rg,  Martin:2004pm, Bozza:2005qg, McFadden:2005mq, Alexander:2007zm},
and the scale-invariant growing mode $\Phi$ can be mapped to the constant mode $\zeta$ in the expanding phase. (It has even been shown that certain types of bounce can process the primordial power spectra in a scale-dependent fashion \cite{Martin:2002ar,Martin:2003bp,Martin:2003sf}.) Furthermore, the predictions of this class of models with respect to the primordial tensor/scalar ratio are also dependent on the details of mode mixing at the bounce, since the ratio $r_\Phi$ (\ref{eq:rPhi}) is generically strongly suppressed, but the ratio $r_\zeta$ (\ref{eq:rzeta}) can be observably large. Therefore, it is necessary to specify in detail the physics of the bounce in order to translate the primordial power spectra generated in the contracting phase to their counterparts in the expanding phase. It remains a tantalizing prospect that bouncing cosmologies can provide a fully viable alternative to inflation.

\section*{Acknowledgments}

WHK is grateful to the Harish Chandra Research Institute in Allahabad, India for hospitality while part of this work was being completed, and thanks L. Sriramkumar for helpful discussions. This research is supported  in part by the National Science Foundation under grant NSF-PHY-0757693.

\end{document}